\documentstyle[prb,aps,preprint]{revtex}
\tighten
\begin{document}
\draft
\title{Spin-flip scattering in the quantum Hall regime}
\author{D. G. Polyakov\cite{DGP}}
\address{Institut f\"ur Theoretische Physik, Universit\"at zu K\"oln, 
Z\"ulpicher Str. 77, 50937 K\"oln, Germany}
\maketitle
\begin{abstract}
We present a microscopic theory of spin-orbit coupling in the integer
quantum Hall regime. The spin-orbit scattering length is evaluated in
the limit of long-range random potential. The spin-flip rate is shown
to be determined by rare fluctuations of anomalously high electric
field. A mechanism of strong spin-orbit scattering associated with
exchange-induced spontaneous spin-polarization is suggested. Scaling
of the spin-splitting of the delocalization transition with the
strength of spin-orbit and exchange interactions is also discussed.
\end{abstract}
\narrowtext

\section{Introduction}

Though the underlying physics of the quantum Hall effect (QHE), both
integer and fractional, can be understood in terms of spinless
electrons, phenomena associated with the spin degree of freedom have
increasingly become a subject of great interest. Essential to their
description are two circumstances: high degeneracy of Landau levels,
peculiar to the QHE regime, and very small Zeeman splitting,
characteristic to semiconductors like GaAs. As a result, mixing of
states with different Landau level numbers $N$ both by disorder and
electron-electron interactions may be weak, while two Zeeman levels
with the same $N$ are strongly overlapped. It is this case that we are
concerned with in the present paper. However, even within a single
Landau level, interplay of disorder and interactions has been proven
to be a very rich subject. Much of the recent interest in the role of
spin in QHE was sparked off by the realization that clean
two-dimensional electron systems in the limit of strong magnetic field
may exhibit novel liquid states with long-range spin correlations.
\cite{girvin95} These developments underline the role of
interactions. With increasing disorder, the number of conducting
phases which arise in succession as the Fermi level sweeps through a
given Landau level decreases, \cite{kivelson92,halperin93} so that
eventually a crossover to the regime of the integer QHE occurs. For
spinless electrons, the latter is characterized by the existence of
only one extended state within the Landau level. We are primarily
interested here in this case. The localization length $\xi(E)$ then
diverges as the energy $E$ approaches the critical value $E_c$
according to $\xi(E)\propto |E-E_c|^{-\gamma}$. \cite{huckestein95}
The importance of spin in this picture has been under much discussion
\cite{khmelnitskii92,lee94a,wang94,hikami93,lee94b,polyakov95,hanna95}
largely because the inclusion of spin splits the delocalization
transition. It means that two extended states with different energies
$E_c^{\pm}$ appear, corresponding to two spin projections. Of course,
if the splitting had merely reflected the fact that there is a finite
bare Zeeman energy, it would be a trivial generalization of the
spinless case: two systems of electrons with opposite spin would
remain independent of each other. The point is that turning on a
spin-orbit (SO) interaction drives these initially uncoupled systems
into a new quantum Hall phase with an internal degree of freedom.

The influence of the SO coupling is in fact twofold. Firstly, it
renormalizes the effective $g$-factor, thus leading to the existence
of a finite Zeeman energy even if the bare $g$-factor is neglected
(quite separate from the exchange enhancement \cite{ando74}). This
naturally yields the splitting of the delocalization transition. The
appearance of two distinct critical energies has been observed by
numerical simulations of spin-degenerate electrons.
\cite{lee94a,wang94,hanna95} Secondly, the SO scattering gives rise to
a random coupling of states with opposite spin. A key issue then is
how the SO coupling affects the critical behavior of the localization
length. Numerical calculations \cite{lee94a,wang94,hanna95} support
the conclusion that, in close vicinity of the critical points,
$\xi(E)$ diverges at $E\to E_c^{\pm}$ with the same critical exponent
$\gamma$ as for spinless electrons. However, a question how $\xi(E)$
at $E={1\over 2}(E_c^++E_c^-)$ scales with the strength of the SO
coupling should be addressed in order to provide a reliable
explanation of reported anomalies \cite{hwang93,koch91a,engel93,wei94}
in the critical broadening of $\sigma_{xx}$-peaks. As argued in Ref.\
5, in the case of short-range disorder $\xi(E)$ in the middle between
two delocalized states scales as the SO-scattering length. On the
other hand, according to Ref.\ 10, in the limit of long-range disorder
the quantum localization length at $E$ lying between $E_c^+$ and
$E_c^-$ is strongly increased due to the SO coupling. In both cases
\cite{khmelnitskii92,polyakov95} the SO interaction does not affect
behavior of $\xi(E)$ at $|E-E_c^\pm|\gg |E_c^+-E_c^-|$.

The crucial parameter which governs behavior of the spin-degenerate
electron system in the quantum Hall regime is the ratio
$L_{so}/\xi(E)$, where $L_{so}$ is the SO-scattering length. The
purpose of this paper is to evaluate $L_{so}$ in the case of a
long-range random potential $V(\bbox{\rho})$. We make essential use of
the adiabatic approximation, which is only accurate in the extreme of
smooth disorder. However, it is this case that is experimentally
relevant both in quantum wires and bulk heterostructures with a large
spacer. Furthermore, it has been shown \cite{polyakov95} that the SO
coupling of percolating trajectories in the limit of long-range
disorder may lead to striking effects in the conductivity at finite
temperature, the effects being stronger the smaller $L_{so}/\xi(E)$ at
the Fermi level.  With this motivation we take the quasiclassical
approach in which the spin-flip scattering occurs when electrons move
along equipotentials of $V(\bbox{\rho})$. Since SO transitions
necessarily involve momentum transfer to the fluctuating potential,
their rate must be suppressed in the limit of smooth disorder. Hence,
the spin-flip scattering in that case must be very sensitive to the
local configuration of the random potential. A question then arises as
to the nature of the SO coupling between two long trajectories with
opposite spin. More specifically, the problem is how to average the
local SO coupling. We show that transitions take place in rare
fluctuations of a ``specific'' shape. These optimum configurations
provide the highest averaged scattering rate, which is much larger
than that obtained \cite{khaetskii92} within the Born approximation. A
second issue of interest is how the gap $|E_c^+-E_c^-|$ depends on the
SO-interaction constant. We obtain different contributions to the
value of the gap and argue that, in general, it does not scale as a
sample-averaged spin-splitting. The opposite conclusion has been drawn
\cite{hanna95} from numerical simulations assuming that fluctuations
of the local SO coupling are uncorrelated with those of
$V(\bbox{\rho})$. Their correlation, however, is important in the case
of smooth disorder.

We first deal with non-interacting electrons. This is a marginal case
even in the integer quantum Hall regime, yet its consideration enables
us to get an insight into the problem with electron-electron
interactions included. As a matter of fact, localization properties of
interacting electrons change strongly with increasing correlation
radius of disorder. If the random potential is short-ranged, the
physics of interaction in the integer QHE is to a great extent
captured by the concept\cite{polyakov93} of the Coulomb
gap,\cite{efros85} according to which the single-particle density of
states at the Fermi level $g(E_F)$ vanishes however small $|E_F-E_c|$
is. The underpinning of this picture is the notion that, whatever
$E_F$, the Coulomb energy on the scale of the one-electron
localization length $\xi(E_F)$ is of the order of the characteristic
energy spacing $\delta_c\sim 1/g(E_F+\delta_c)\xi^2(E_F)$ on the same
scale. This naturally implies that electron-electron interactions
cannot affect the critical behavior of the localization length at the
Fermi level. This same conclusion can be reached on more
phenomenological grounds if the dynamic scaling exponent $z$ is set to
1 (to put it another way, dynamics of the wave packet on scales
shorter than $\xi(E_F)$ must be governed by charge spreading according
to the Ohm's law rather than slower diffusion; accordingly,
$g(E_F+\delta_c)\sim 1/\hbar D(\xi(E_F))$, where the effective,
scale-dependent diffusion coefficient $D(\xi)\propto \xi$ is
introduced).\cite{fisher90,lee95} The suppression of $g(E_F)$ has been
confirmed numerically. \cite{yang93} Experimental data \cite{koch95}
on $\sigma_{xx}$-peak broadening agree well with the Coulomb gap
approach. \cite{polyakov93} Recently, the scaling behavior of
$\xi(E_F)$ at $E_F\to E_c$ has been observed directly by numerical
simulation within the Hartree-Fock scheme. \cite{yang95} Thus, it is
the strong effect on the density of states that is the reason for no
effect on the localization length. Clearly, inherent to the formation
of the sharp Coulomb gap is the absence of screening at large scales,
in the sense that in the Coulomb glass system \cite{efros85} the
interaction between two localized states at large distances $\rho$
behaves as $\rho^{-1}$. In the limit of smooth random potential,
however, dielectric properties of disordered electrons become
essentially different. Screening then is strong and takes place on
scales shorter than the correlation radius of the potential. In the
quantum Hall regime with smooth disorder, a finite screening radius is
due to electron-electron correlations \cite{efros88a} and, naturally,
it cannot be smaller than the magnetic length. As a result of the
non-perfect screening, there exists a random self-consistent
potential, whose fluctuations in the extreme of high field are
correlated on the same scale as for the bare potential. It is likely
that at zero temperature the picture of percolation through this
potential is much the same as commonly used within the one-electron
consideration, which implies a sharp Fermi distribution for
percolating particles. It is expected also
\cite{kivelson92,halperin93} that with smoothing disorder a series of
delocalization transitions between fractional Hall phases
occurs. However, given the fact that within the electrostatic approach
\cite{luryi87,efros88a,chklovskii92} the percolating trajectory is
partially occupied in a finite range of filling factor, there remains
a challenging question, even within the integer quantum Hall regime,
about evolution of the phase diagram of the interacting electron
system as the correlation radius of disorder is increased.

Having got the aforesaid results for $L_{so}^{-1}$ and $|E_c^+-E_c^-|$
within the one-particle picture, we proceed to include effects of
electron-electron interactions on these two quantities. In the case of
long-range disorder, turning on interactions has two immediate
consequences, which are screening and exchange-induced enhancement of
Zeeman splitting. \cite{ando74} Evidently, both effects yield
suppression of the SO scattering. Thus, the gain in the spin-flip rate
we obtain by finding that actually it is determined by optimum
fluctuations seems to be lost once the interactions are taken into
account. In fact, however, interplay of the direct and exchange
interactions leads to the appearance of a specific mechanism of strong
SO scattering. Recently, a great deal of attention has been given to
understanding the structure of edge channels in a quantum wire in a
crossover region between regimes of strong and weak, as compared with
the strength of electron-electron interactions, confinement.
\cite{dempsey93,kinaret90,dechamon94,manolescu95,fogler95} It has been
shown, \cite{dempsey93,kinaret90} within the Hartree-Fock approach,
that these two limiting cases are separated by a phase transition at
which the spontaneous spin-polarization occurs with decreasing slope
of confinement. With further advancing into the interaction-dominated
regime, regions occupied by compressible liquid appear.
\cite{dempsey93,dechamon94} In the disordered system at issue, the
critical points, at which the separation between two edges with
opposite spin changes in a sharp manner, are randomly distributed
along electron trajectories. It is the series of phase transitions at
these points with which we associate the enhancement of scattering,
since the sharp change of the effective scattering potential favors
the SO transitions. As for the influence of electron-electron
interactions on the gap $|E_c^+-E_c^-|$, we argue that this quantity
does not exhibit critical behavior when the area occupied by
compressible liquid becomes comparable to that with integer filling,
which might be expected within the framework of the mean-field theory
developed in Ref.\ 32 for large $N$. We demonstrate that the
spontaneous spin-polarization which inevitably occurs at the critical
saddle points of the percolation network gives rise to a finite
splitting of the metal-insulator transition even in the
disorder-dominated regime.

In accordance with our approach outlined above, the body of the paper
consists of two following parts. After formulating the basics of the
SO coupling in the presence of a magnetic field, we derive the
spin-flip length $L_{so}$ within the one-electron picture. We discuss
also the effect of the SO coupling on the temperature broadening of
$\sigma_{xx}$ peaks, as well as on the splitting of the delocalization
transition. In the second part, we discuss effects of
electron-electron interactions and interplay of the interactions and
the SO coupling.

\section{Spin-orbit coupling of Landau level states}

The Hamiltonian of a 2D electron contains the term 
\begin{equation}
{\cal H}_{so}(\bf k)=\alpha{\bf n}(\bbox{\sigma}\times{\bf k})~,
\end{equation} 
which couples the spin $\bbox\sigma$ to the kinetic momentum \(\hbar
{\bf k}=-i\hbar\frac{\partial}{\partial\mbox{\boldmath $\rho
$}}+\frac{e}{c}\bf A\) even in a homogeneous 2D system.
\cite{bychkov84} The SO term originates from the asymmetry of the
confining potential and the constant $\alpha$ is a measure of the
asymmetry, ${\bf n}$ being the unit vector along the normal to the 2D
plane. In GaAs-type cubic crystals, which lack inversion symmetry, the
2D Hamiltonian of a free electron has yet another SO term linear in
$\bf k$ which persists even in a symmetric confinement.
\cite{dyakonov86} This depends on the orientation of $\bf n$ with
respect to the crystal axes and has exactly the same structure as the
term (1) if ${\bf n}\parallel [111]$ but, say, is proportional to
${\bf n}(\bbox{\tilde\sigma}\times{\bf k})$, where
$\tilde\sigma_x=\sigma_y$, $\tilde\sigma_y=\sigma_x$ with
$\bbox{\hat{x}}$ and $\bbox{\hat{y}}$ along the principal directions
and $\bbox{\hat{z}}$ along $\bf n$, if ${\bf n}\parallel [001]$. For
the sake of simplicity, the SO interaction is treated here in the
isotropic form (1) -- taking the anisotropy into account will merely
renormalize the constant $\alpha$ in the final expressions for the
SO-scattering length.

To begin with, let us find how the Landau level states are modified by
the inclusion of the SO interaction (1) in the presence of a
homogeneous in-plane electric field $\bbox{\cal E}_0$. The Hamiltonian
reads ${\cal H}={\cal H}_0+{\cal H}_{so}$, where ${\cal
H}_0={\hbar^2{\bf k}^2\over 2m}+e\bbox{{\cal
E}}_0\bbox{\rho}-\Delta_0\sigma_z$, $2\Delta_0$ being the bare Zeeman
splitting. It is convenient to represent ${\cal H}_{so}$ as a sum of
two terms, ${\cal H}_{so}^{(c)}=\alpha{\bf n}(\bbox{\sigma}\times({\bf
k}-{m\over\hbar}{\bf v}_d))$ and ${\cal H}_{so}^{(d)}=-{\alpha
e\over\hbar\omega_c}(\bbox{\sigma}\cdot\bbox{\cal E}_0)$, where ${\bf
v}_d$ is the drift velocity and $\omega_c$ the cyclotron frequency.
The first term couples the spin to the cyclotron rotation whereas the
latter yields coupling of the spin to the drift motion.  ${\cal
H}_{so}^{(c)}$ has only nonzero matrix elements between the states of
the Landau levels $N$ and $N\pm 1$ while ${\cal H}_{so}^{(d)}$ is
diagonal in $N$. Taking ${\bf A}=Bx\bbox{\hat{y}}$ and $\bbox{{\cal
E}}_0=-{\cal E}_0\bbox{\hat{x}}$, we write the eigenfunctions of
${\cal H}_0$ with the energies
$\epsilon^{\pm}_{Nq}=(N+\frac{1}{2})\hbar\omega_c+e{\cal
E}_0q\lambda^2\mp\Delta_0-{e^2{\cal E}^2_0\over 2m\omega_c^2}$ as
$\Psi_{Nq}^{\pm}=\exp (iqy)\psi_{Nq}^{\pm}(x+q\lambda^2-{e{\cal
E}_0\over m\omega_c^2})$. Inclusion of ${\cal H}_{so}^{(c)}$ couples
the lower Zeeman state of the level $N$ to the upper state of the
level $N+1$ and the upper Zeeman state $N$ to the lower state $N-1$.
The modified eigenfunctions take the form
\begin{equation}
  \tilde\Psi_{Nq}^{\pm}=\Psi_{Nq}^{\pm}\mp\vartheta^{\pm}_N\Psi_{N\pm
    1,q}^{\mp}~,\quad \vartheta^+_N=\vartheta^-_{N+1}~.
\end{equation} 
We suppose the SO interaction to be weak in the sense that
$\vartheta^{\pm}_N\ll 1$, in which case $\vartheta^-_N\simeq
{\alpha\sqrt{2N}\over \lambda\hbar\omega_c}$ and the eigenenergies
$\tilde E^{\pm}_{Nq}\simeq
E^{\pm}_{Nq}\mp(\vartheta^{\pm}_N)^2\hbar\omega_c$. It follows that
${\cal H}_{so}^{(c)}$ increases the Zeeman splitting, which becomes
equal to $2(\Delta_0+\Delta_c)$, where $\Delta_c =
[(\vartheta^-_N)^2+(\vartheta^+_N)^2]\hbar\omega_c/2=
(2N+1)m\alpha^2/\hbar^2$. Taking now ${\cal H}_{so}^{(d)}$ into
account causes coupling of the modified states $\tilde\Psi_{Nq}^{\pm}$
with different spin projections within the same Landau level. The wave
functions are finally given by $(\sigma=\pm)$
\begin{equation}
  \Phi^\sigma_{Nq}=\sum_{\sigma^\prime}\Omega_{\sigma\sigma^\prime}
  (\theta_N)\tilde\Psi_{Nq}^{\sigma^\prime}~,\quad
  \hat{\Omega}(\theta_N)=\left( \begin{array}{rr} \cos{\theta_N\over
  2} & -\sin{\theta_N\over 2} \\ \sin{\theta_N\over 2} &
  \cos{\theta_N\over 2}
\end{array} \right)~,
\end{equation} 
where the matrix $\hat{\Omega}(\theta_N)$ expresses a rotation through
the angle $\theta_N$ about the $y$ axis in spinor space. The rotation
angle is defined by $\sin\theta_N=\Delta_d/\Delta$, where
\begin{equation} 
  \Delta=\sqrt{(\Delta_0+\Delta_c)^2+\Delta_d^2}~,\quad
  \Delta_d={\alpha e{\cal E}_0\over \hbar\omega_c}~.
\end{equation}
In the high-$\cal E$ limit $\theta_N\to \pi/2$. The quantity
$2\Delta$, given in Eq.\ (4), is the Zeeman splitting renormalized by
the SO interaction in the homogeneous field $\bbox{\cal E}_0$. Eq.\
(4) is valid so long as $\Delta\ll\hbar\omega_c$; the splitting due to
$\Delta_d$ and $\Delta_c$ was derived in Refs.\ 17 and 11,
respectively.

Now let us take into consideration a random scalar potential
$V(\bbox{\rho})$, which, in the presence of the SO coupling (1), gives
rise to spin-flip processes.  We neglect the random-potential-induced
mixing of states with different $N$ and seek a solution of the
equation $({\cal H}_0+{\cal H}_{so}+V-E)\Phi_N=0$ of the form
\begin{equation} 
  \Phi_N=\sum_\sigma\int {dq\over 2\pi} C^\sigma_{Nq}\Phi^\sigma_{Nq}
\end{equation}
with $\Phi_{Nq}^\sigma$ defined in Eqs.\ (2,3).  The amplitudes
$C^\sigma_{Nq}$ then satisfy the integral equations
\begin{equation}
  (E^\sigma_{Nq}-E)C^\sigma_{Nq}+\sum_{\sigma^\prime}\int
  {dq^\prime\over 2\pi}U_{\sigma\sigma^\prime}(q,q^\prime)
  C^{\sigma^\prime}_{Nq^\prime}=0~, \quad
  U_{\sigma\sigma^\prime}(q,q^\prime)=\int
  d\bbox{\rho}V(\bbox{\rho})(\Phi_{Nq}^\sigma)^*\Phi_{Nq^\prime}^
  {\sigma^\prime}~,
\end{equation}
where $E_{nq}^\sigma$ are the eigenenergies of ${\cal H}_0+{\cal
H}_{so}$. The matrix elements $U_{\sigma\sigma^\prime}$ can be written
in the limit $\vartheta_N^\pm\ll 1$ as
\begin{eqnarray}
  && U_{++}(q,q^\prime)=U_{--}(q,q^\prime)=\int{dk\over
    2\pi}V_{k,q-q^\prime}e^{-ik\left( {q+q^\prime\over
    2}\lambda^2-{e{\cal E}_0\over m\omega_c^2}\right)}I_N\left(
    [k^2+(q-q^\prime)^2]{\lambda^2\over 2}\right)~,\\ &&
    U_{+-}(q,q^\prime)=U_{-+}^*(q^\prime,q)= \nonumber \\ &&
    -{\alpha\over \hbar\omega_c}\int{dk\over 2\pi}[ik\cos\theta_N +
    (q-q^\prime)]V_{k,q-q^\prime}e^{-ik\left( {q+q^\prime\over
    2}\lambda^2-{e{\cal E}_0\over m\omega_c^2}\right)}I_N\left(
    [k^2+(q-q^\prime)^2]{\lambda^2\over 2}\right)~
\end{eqnarray}
with $V_{k_x,k_y}=\int d\bbox{\rho}V(\bbox{\rho})\exp
(-i\bbox{k\rho})$ and $I_N(t)=\exp (-{t\over 2})L_N(t)$, $L_N$ being
the Laguerre polynomial.  The diagonal matrix elements (7) are
responsible for the curvature of quasiclassical trajectories in the
presence of $V(\bbox{\rho})$, which otherwise are straight lines. The
matrix elements (8) with different spin-projections yield the SO
scattering (as well as a renormalization of the local Zeeman splitting
by the electric field \({1\over e}{dV\over d\mbox{\boldmath
$\rho$}}\)).  They are nonzero only because of the SO admixture of
higher-Landau-level states [Eq.\ (2)].

Consider first a single-scattering problem: Given a potential
$V(\bbox{\rho})$ which vanishes at infinity and an electron in the
state $\Phi^-_{Nq_-}$ incident on the scattering region from
$y=-\infty$, find the spin-flip amplitude $t_{-+}$ in the asymptotic
form $t_{--}\Phi^-_{Nq_-}+t_{-+}\Phi^+_{Nq_+}$ of the transmitted wave
at $y\to\infty$. Here the wave vectors $q_\sigma$ are defined
according to $E^\sigma_{Nq_\sigma}=E$. Since $\partial
E^\sigma_{Nq}/\partial q = e{\cal E}_0\lambda^2$, the first-order
result is readily obtained to be
\begin{equation}
  t^{(1)}_{-+}=\int{dq\over 2\pi}{U_{+-}(q,q_-)\over
    E-E^+_{Nq}}=-{i\over e{\cal E}_0\lambda^2}U_{+-}(q_+,q_-)~.
\end{equation}
It is valid if the external field ${\cal E}_0$ much exceeds the
characteristic scattering field. To go beyond the perturbation
picture, we make use of the assumption that the characteristic radius
of the scattering potential $d\gg \lambda\sqrt{N+1}$ and, introducing
the one-dimensional functions $G_N^\sigma(y)=\int{dq\over
2\pi}C^\sigma_{Nq}\exp (iqy)$, seek them in the quasiclassical form
\begin{equation}
  G^\sigma(y)=\sum_{\sigma^\prime}g_{\sigma\sigma^\prime}(y)
  e^{iS_{\sigma^\prime}(y)}
\end{equation}
(we drop the index $N$ from now on), where $S_\sigma$ and
$g_{\sigma\sigma^\prime}$ are smooth functions on the scale of
$\lambda\sqrt{N+1}$. Besides, we suppose the characteristic distance
between quasiclassical trajectories with opposite spin to be much
smaller than $\lambda$. This allows us to replace $I_N$ by 1 in Eqs.\
(7,8). Substituting the expression (10) into the equations
\begin{equation}
  -e{\cal E}\lambda^2\left(q_\sigma+i{d\over
      dy}\right)G^\sigma(y)+\sum_{\sigma^\prime}\int dy^\prime
      u_{\sigma\sigma^\prime}(y,y^\prime)G^{\sigma^\prime}(y^\prime)=0~,
\end{equation}
where $u_{\sigma\sigma^\prime}(y,y^\prime)$ are the Fourier components
of $U_{\sigma\sigma^\prime}(q,q^\prime)$, we expand
$S_\sigma=S^{(0)}_\sigma+S^{(1)}_\sigma+\ldots$ in powers of
$(\lambda/d)^2$ and get, in the first approximation,
\begin{equation}
  -e{\cal E}_0\lambda^2\left(q_\pm-{dS^{(0)}_\pm\over
      dy}\right)+V\left(x_\pm
      (y),y\right)\mp[\Delta(y)-\Delta]=0~,\quad
      x_\sigma(y)=-{dS^{(0)}_\sigma\over dy}\lambda^2+{e{\cal
      E}_0\over m\omega_c^2}~.
\end{equation}
Here $x_\sigma(y)$ are in fact the quasiclassical equations of motion
for electrons with opposite spin. The function $2\Delta(y)$ defines
the local Zeeman splitting
$2[(\Delta_0+\Delta_c)^2+\Delta^2_d(y)]^{1/2}$, where $\Delta_d(y)=
\alpha e {\cal E}(y)/\hbar\omega_c$. In this last expression
$\bbox{{\cal E}}(y)$ is the total field an electron experiences while
traveling along the quasiclassical trajectories: $\bbox{{\cal
E}}(y)=\bbox{{\cal E}}_0+\bbox{\varepsilon}(y)$,
\(\bbox{\varepsilon}(y)={1\over e}{\partial V\over \partial
\mbox{\boldmath $\rho$}}|_{x=x(y)}\). It is legitimate to drop the
subscript $\sigma$ in the definition of $\bbox{\cal E}(y)$ since
$|x_+(y)-x_-(y)|\ll d$. The next term in the expansion of $S_\sigma$
(the last one we need to keep) satisfies the equation
\begin{equation}
  -{\cal E}_x(y){dS^{(1)}\over dy}+{i\over 2}{d{\cal E}_x(y)\over
    dy}=0~,
\end{equation}
the same for both $\sigma$, which gives nothing but the
drift-velocity-dependent prefactor $\exp(iS^{(1)})\propto
1/\sqrt{-{\cal E}_x(y)}$.  In the quasiclassical limit, the amplitudes
$g_{\sigma\sigma^\prime}(y)$ obey somewhat cumbersome relations
\begin{equation}
  {g_{-+}\over g_{++}}=-\left({g_{+-}\over
      g_{--}}\right)^*={\tan\left({\theta(y)\over
      2}\right)e^{i\varphi(y)}-\tan\left({\theta\over 2}\right)\over
      1+\tan\left({\theta(y)\over 2}\right)\tan\left({\theta\over
      2}\right)e^{i\varphi(y)}}~,\quad \tan\left({\theta(y)\over
      2}\right)={\Delta_d(y)\over \Delta_0+\Delta_c+\Delta(y)}~,
\end{equation}
where $\varphi(y)$ is the polar angle of the total field $\bbox{\cal
  E}(y)$ and $\theta=\theta(y\to\pm\infty)$. However, one can
  recognize them as the formulae describing an adiabatic rotation of
  the spin caused by the SO interaction with the scattering field
  $\bbox{\varepsilon}(y)$. Coupling between the waves with different
  $S_\sigma$ in the last term of Eq.\ (11), which breaks the
  adiabaticity, gives then for the spin-flip amplitude
  $t_{-+}=g_{-+}(\infty)/g_{--}(-\infty)$:
\begin{equation}
  t_{-+}= -i{\alpha m\over \hbar^2}\int{dy\over {\cal
      E}_x}\left(\chi^+\right)^*(\bbox{\sigma{\cal
      E}})\chi^-e^{i(S_--S_+)}~,
\end{equation}
where $\chi^\sigma$ are the spinors subjected to the rotation
specified by the angles $\theta(y)$ and $\varphi(y)$. The difference
$\phi(y)=S_-(y)-S_+(y)$ is identified with the number of magnetic flux
quanta in the area enclosed between two quasiclassical trajectories
with opposite spin. Since the distance between the trajectories is
much smaller than $d$, $\phi(y)$ can be conveniently written in the
form
\begin{equation}
  \phi(y)=-{2\over e\lambda^2}\int_{0}^{y}dy^\prime{\Delta(y^\prime)
    \over {\cal E}_x(y^\prime)}~.
\end{equation}
Upon using the drift-motion equation ${dl\over dy}=-{{\cal E}(l)\over
  {\cal E}_x(l)}$, where $l$ is the longitudinal coordinate along the
  trajectory, the spin-flip amplitude can be re-expressed in terms of
  the integral along the classical path:
\begin{equation}
  t_{-+}={i\over \hbar}{\alpha e\over \hbar\omega_c}\int {dl\over
    v_d(l)}\left(\chi^+\right)^*\left(\bbox{\sigma{\cal
    E}}(l)\right)\chi^-e^{i\phi(l)}~,
\end{equation}
$v_d(l)$ being the drift velocity at the point $l$ of the
trajectory. We infer from this expression that the operator
\begin{equation} 
  \tilde{{\cal H}}_{so}=-{\alpha e\over
    \hbar\omega_c}(\bbox{\sigma{\cal E}})~,
\end{equation}
plays the role of the scattering potential responsible for the SO
transitions. Unlike the ``conventional" SO term, which is proportional
to $\bbox{\sigma}({\bf k}\times \bbox{\cal E})$, the potential (18)
leads to the relaxation of the spin polarized along the normal to the
2D plane.  Moreover, being proportional to the built-in electric field
(hidden in the constant $\alpha$), it is typically much stronger. It
is worthwhile to notice that, though $\tilde{{\cal H}}_{so}$ yields
the spin-flip scattering within the same Landau level, in the course
of its derivation we had to take into account mixing of the Landau
level states with different $N$ [Eq.\ (2)]. It may be viewed as a
projection of ${\cal H}_{so}$ [Eq.\ (1)] onto a given Landau level in
the presence of smooth disorder. We observe also that provided
$\chi^\sigma$ are the eigenfunctions of $\sigma_z$, which is the case
at $\Delta(y)\gg\Delta_d(y)$, only the phase factors remain in the
integrand of Eq.\ (17):
\begin{equation}
  t_{-+}=-i{\alpha m\over \hbar^2}\int dy\left[ 1-i{{\cal E}_y(y)\over
      {\cal E}_x(y)}\right] e^{i\phi(y)}=i{\alpha m\over \hbar^2}\int
      dle^{i[\phi(l)-\varphi(l)]}~.
\end{equation}

\section{Spin-flip scattering length}

Now we can turn to evaluation of the SO scattering length $L_{so}$,
defined according to
$L_{so}^{-1}=\lim_{L\to\infty}L^{-1}<|t_{-+}|^2>$, where $L$ is the
trajectory length and $<>$ denotes averaging over fluctuations of the
electric field $\bbox{\cal E}(\bbox{\rho})$. For definiteness, let the
fluctuating part of the field, $\bbox{\varepsilon}(\bbox{\rho})$, be
created by ionized impurities randomly distributed with a sheet
density $n_i$ in a thin layer separated from the electron gas by an
undoped spacer of the width $d\gg n_i^{-1/2}$. Since we deal with the
case of a long-range random potential, $d$ must be much larger than
$\lambda$. The correlation function of the fluctuations reads
$<\bbox{\varepsilon\varepsilon}>_{\bf k}=8\pi{\cal E}_t^2d^2e^{-2kd}$,
${\cal E}_t^2={\pi\over 2}{e^2n_i\over \epsilon^2 d^2}$, $\epsilon$
being the dielectric constant. Suppose first that the homogeneous
field ${\cal E}_0\gg {\cal E}_t$, so that typically the fluctuations
only slightly curve the quasiclassical trajectories.  In addition, let
us focus on the case $\Delta\gg \Delta_d$.  Then, in the first
approximation,\cite{khaetskii92} equivalent to Eq.\ (9), the curving
of the trajectories can be ignored in the argument of
$\bbox{\varepsilon}(y)$ and one can set ${\cal E}_x(y)\to -{\cal E}_0$
in Eqs.\ (15,16), which gives \begin{equation}
  {1\over L_{so}^{(1)}}=\left({\alpha m\over \hbar^2{\cal
        E}_0}\right)^2\int_{-\infty}^{\infty} dy {\cal K}_{yy}(y)\exp
        \left(i{{\cal E}_\Delta\over {\cal E}_0}{y\over
        2d}\right)=4\left({\alpha m\over \hbar^2}\right)^2d{{\cal
        E}_\Delta {\cal E}_t^2\over {\cal E}_0^3} K_1\left({{\cal
        E}_\Delta\over {\cal E}_0}\right)~.
\end{equation}
In the above, ${\cal
  K}_{ii}(y)=<\varepsilon_i(0,0)\varepsilon_i(0,y)>$, ${\cal
  K}_{yy}(y)= {\cal E}_t^2[1+(y^2/4d^2)]^{-3/2}-{\cal K}_{xx}(y)$,
  ${\cal K}_{xx}(y)={\cal E}^2_t(4d^2/y^2)[1+
  (y^2/4d^2)]^{-1/2}\{[1+(y^2/4d^2)]^{1/2}-1\}$, ${\cal
  E}_\Delta=4\Delta d/ e\lambda^2$, and $K_1$ is the Bessel function.
  The characteristic parameter ${\cal E}_\Delta/{\cal E}_0$, appeared
  in Eq.\ (20), is represented as ${\cal S}/\lambda^2$, $\cal S$ being
  the area between the trajectories with different spin on the scale
  of $2d$. The spin-flip rate falls off rapidly with growing this
  parameter, as $\exp (-{\cal E}_\Delta/{\cal E}_0)$, when ${\cal
  E}_\Delta\agt{\cal E}_0$. This simply reflects the fact that
  scattering with large momentum transfer is suppressed in the limit
  of smooth disorder.  Let us show, however, that in this limit the
  fluctuations of the phase $\phi(y)$ lead to a drastic enhancement of
  the scattering rate in comparison with Eq.\ (20). First of all,
  notice that the curvature of the trajectories in Eq.\ (16) should be
  taken into account if the {\it rms} fluctuation of the area
  $<(\delta {\cal S})^2>^{1/2}\sim \left({\cal E}_\Delta{\cal E}_t/
  {\cal E}_0^2\right)\lambda^2$ exceeds $\lambda^2$, i.e., if ${\cal
  E}_\Delta/{\cal E}_0\agt {\cal E}_0/{\cal E}_t$. Provided ${\cal
  E}_\Delta\gg{\cal E}_0$, this occurs when ${\cal E}_t$ is still much
  smaller than ${\cal E}_0$, and the trajectories are almost straight
  lines. Expanding $\phi(y)$ to first order in $\varepsilon_x$ gives
  then
\begin{equation} 
  \left< e^{i\phi(y)}\right>\to \exp \left(i{{\cal E}_\Delta\over
      {\cal E}_0}{y\over 2d}-{{\cal E}_\Delta^2\over 8{\cal
      E}_0^4d^2}\int_{0}^{y}dy^\prime\int_{0}^{y}dy'' {\cal
      K}_{xx}(y'-y'')\right)~.
\end{equation} 
Owing to the fast oscillations associated with the first term in the
exponent, the second term, which describes the decay of the average
$<\exp(i\phi)>$ with increasing disorder, can simply be picked up at
the singular point of the pre-exponential ${\cal K}_{yy}(y)$ in the
upper half-plane, $y=2id$ (then the integration in Eq.\ (21) should be
effected along the straight line connecting the points $y=0$ and
$y=2id$). The result is
\begin{equation}
  {1\over L_{so}}={1\over L_{so}^{(1)}}\exp \left[ (1-\ln 2){{\cal
        E}_\Delta^2{\cal E}_t^2\over {\cal E}_0^4}\right]~,\quad
        {1\over L_{so}^{(1)}}= 2\sqrt{2\pi}\left( {\alpha m\over
        \hbar^2}\right)^2 d{{\cal E}_t^2\over {\cal
        E}_0^2}\sqrt{{{\cal E}_\Delta\over {\cal E}_0}}\exp \left(
        -{{\cal E}_\Delta\over {\cal E}_0}\right)~.
\end{equation}
The additional exponential factor $L_{so}^{(1)}/L_{so}$ is much larger
than unity if ${\cal E}_\Delta/{\cal E}_0\gg {\cal E}_0/{\cal E}_t$,
though the relative correction to the exponent of the spin-flip rate
remains small as long as ${\cal E}_\Delta/{\cal E}_0\ll {\cal
E}_0^2/{\cal E}_t^2$. In fact, Eq.\ (22) signals that at ${\cal
E}_\Delta/{\cal E}_0\agt {\cal E}_0^2/{\cal E}_t^2$ a new regime of
the SO scattering should appear, which is dominated by the
fluctuations of $\phi(y)$. At the crossover point to this regime, the
ratio ${\cal E}_t/{\cal E}_0$ is still small, $\sim ({\cal E}_0/{\cal
E}_\Delta)^{1/2}$.  Nevertheless, the expansion (21) in powers of
$\varepsilon_x/{\cal E}_0$ under the sign of averaging is not valid
any more, for the scattering rate at ${\cal E}_\Delta/{\cal E}_0\gg
{\cal E}_0^2/{\cal E}_t^2$ is determined by rare fluctuations in which
the local electric field greatly exceeds the homogeneous component
${\cal E}_0$. This can be seen by writing the exponent of
$L_{so}^{-1}\propto \exp (-W)$ as a sum of two terms:
\begin{equation}
  W=W_1+W_2~,\quad W_1={1\over 2}\int {d{\bf k}\over
(2\pi)^2}{|\varepsilon_{x{\bf k}}^{(o)}|^2\over
<\varepsilon_x\varepsilon_x>_{\bf k}}~,\quad W_2=i{2\Delta\over e
\lambda^2}\int_{0}^{\pm 2id}{dy\over -{\cal
E}_0+\varepsilon_x^{(o)}(0,y)}~,
\end{equation}
where $W_1$ represents the probability for the optimum fluctuation
$\bbox{\varepsilon}^{(o)}(\bbox{\rho})$ to occur and $W_2$ stands for
the probability of the spin-flip scattering in the field of this
fluctuation. We have for convenience written $W_1$ in the form of a
functional of $\varepsilon_{x{\bf k}}^{(o)}=\int d\bbox{\rho} \exp
(-i\bbox{k\rho})\varepsilon_x^{(o)}(\bbox{\rho})$, since $W_2$
involves only the $x$-component of the electric field.  Then the
optimum fluctuation satisfies the variational equation $\delta
W/\delta\varepsilon_x^{(o)}=0$ without any subsidiary conditions. The
form of $W_2$ results from the fact that the exponent of the
scattering probability is determined by the phase $\phi(y)$ taken at
the singular points of the sought function
$\bbox{\varepsilon}^{(o)}(\bbox{\rho})$ (upon substitution $x=x(y)$).
These must coincide with the singular points of the correlator
$<\bbox{\varepsilon}(0)\bbox{\varepsilon}(\bbox{\rho})>$, which are
$y=\pm 2id$ at $|x(y)-x(0)|\ll d$. The sign of the upper limit of
integration in $W_2$ should be so chosen that $W_2$ is positive.  If
$W_2$ be expanded as a series in $\varepsilon_x/{\cal E}_0$, keeping
only the first-order term reproduces the exponent of $L_{so}^{-1}$ in
Eq.\ (22). Now, by contrast, let us neglect ${\cal E}_0$ in the
denominator of $W_2$. Then the evaluation of the first variation of
$W$ yields the relation (for $t$ in the interval $(0,2d)$)
\begin{equation} 
\varepsilon_x(x,it)= -{2\Delta\over e\lambda^2}\int_{0}^{2d}{dt'\over
\left[\varepsilon_x(0,it')\right]^2}<
\varepsilon_x(0,0)\varepsilon_x(x,i(t-t'))>~,
\end{equation}
which, by means of analytic continuation, defines the optimum
fluctuation. Dimensional analysis of the equation shows that the
scaling form of the solution is
$\varepsilon_x^{(o)}(\bbox{\rho})={\cal E}_{opt}f(\bbox{\rho}/d)$,
${\cal E}_{opt}={\cal E}_\Delta^{1/3}{\cal E}_t^{2/3}$, $f$ being a
dimensionless function. It follows that the homogeneous field ${\cal
E}_0$ indeed plays no role at ${\cal E}_{opt}\gg {\cal E}_0$, even
though ${\cal E}_0\gg {\cal E}_t$.  Thus the spin-flip rate in this
limit is given by (the pre-exponential factor is determined by $|{\cal
E}_y|\sim {\cal E}_t$)
\begin{equation}
  {1\over L_{so}}\sim \left({\alpha m\over \hbar^2}\right)^2 d\exp
\left[ -\eta\left({{\cal E}_\Delta\over {\cal
E}_t}\right)^{2/3}\right]~,
\end{equation}
where $\eta$ is a numerical coefficient of order unity, i.e., $L_{so}$
is much shorter in comparison with Eq.\ (20) at ${\cal E}_\Delta\gg
{\cal E}_t$. If we choose
$\bbox{\varepsilon}^{(o)}(\bbox{\rho})=[2\varepsilon_x^{(o)}(0)/{\cal
E}_t^2]<\varepsilon_x(0)\bbox{\varepsilon}(\bbox{\rho})>$ as a trial
function with the variational parameter $\varepsilon^{(o)}_x(0)$ in
order to minimize $W$, the estimate (and the upper bound on $\eta$)
$\eta\simeq 1.5$ can be readily obtained. To justify the choice, note
that this trial function gives exact solution for the auxiliary
problem of determining the shape of the most probable fluctuation with
a high electric field fixed only at the point $\bbox{\rho}=0$.

Since ${\cal E}_0$ does not figure in Eq.\ (25), the latter actually
gives $L_{so}^{-1}$ in the random system without any external field,
provided ${\cal E}_\Delta\agt {\cal E}_t$. To find the scattering rate
at ${\cal E}_0=0$ in the opposite limit, ${\cal E}_\Delta\ll {\cal
E}_t$, we omit the phase factor $\exp(i\phi)$ in Eq.\ (19). Then
$L_{so}^{-1}$ can be written as
\begin{equation}
  {1\over L_{so}}=\left({\alpha m\over
      \hbar^2}\right)^2\int_{-\infty}^{\infty} dl \left< e^{i\left[
      \varphi(l)-\varphi(0)\right]}\right>\sim \left({\alpha m\over
      \hbar^2}\right)^2d~,
\end{equation}
where the configurational averaging is performed at a fixed path $l$.
Despite having so simple form, the integral cannot be evaluated
analytically, for the average is determined by $l\sim d$ where the
fractal properties of the trajectories become important. It is worth
remarking that the controlling parameter ${\cal E}_\Delta/{\cal E}_t$
(cf.\ Eqs.\ (25,26)) may be represented as the ratio
$\Delta/\Delta_t$, where $\Delta_t=\Gamma (\lambda/d)^2$ is the
characteristic width of the energy band around the level $E=0$ within
which the tunneling through saddle-points of the random potential is
crucial for the localization properties of the electron gas. Whatever
$d$ is, if $|E|\ll\Delta_t$, the localization-length exponent takes
the universal value $\simeq 2.3$ (this limit corresponds to the
network model \cite{chalker88}). For the localization problem, the
classical percolation approach applies only outside the ``tunneling"
band, yet the above consideration of the spin-flip rate is valid at
$|E|\to 0$ as well.  This is because the SO scattering occurs far away
from the critical saddle points (where the tunneling between the
critical percolation clusters and the interference of
multiple-scattered waves take place).  Thus, if $\Delta\alt\Delta_t$,
the spin-flip rate is given by Eq.\ (26) even within the tunneling
band.  At $\Delta\gg \Delta_t$ it falls off with increasing Zeeman
splitting according to Eq.\ (25). Here we must recall, however, that
we have limited ourselves to the case when the distance between two
states with different spin is smaller than $\lambda$.  Otherwise, the
scattering rate would acquire an additional small factor associated
with the weak spatial overlap of the states. It is easy to see that,
for a typical place of the electron trajectories, this restriction
breaks if $\Delta\agt\Gamma\lambda/d$. So the allowed range of
$\Delta$ is much larger than $\Delta_t$. But, in fact, our conclusions
hold in a still wider range of $\Delta$, namely $\Delta\alt
\Gamma(\lambda/d)^{1/2}$, for it is necessary that the distance
between the states be smaller than $\lambda$ only in the regions where
the scattering actually occurs (i.e., where the electric field $\sim
{\cal E}_{opt}$ is much higher than the typical one).

We should note that writing down $W_2$ in the form (23) we in fact
oversimplified the problem. Eq.\ (23) is the contribution to
$L_{so}^{-1}$ which comes in the short-wave limit from the singular
points of the function ${\cal E}_y(y)$. As follows from Eq.\ (19),
these are not the only singular points of the pre-exponential factor
in the integrand of $L_{so}^{-1}$ -- special consideration must be
given to zeros of the function ${\cal E}_x(y)$ in the upper
half-plane.  It is by no means obvious that, as we will see below,
they are of no importance in the problem under
discussion. ``Dangerous" to the evaluation of $L_{so}^{-1}$ are zeros
close to the real axis. We associate them with kinks of the
quasiclassical trajectories which occur, however large $d$ is, in
vicinity of the saddle points of the random potential (apart from the
critical saddle points, which connect up the critical clusters to each
other so as to form the percolation network, there are a lot of other
saddle points which an electron hits on its way between the critical
ones). Though the electric field gets weaker near the saddle points,
its orientation changes sharply there, which favors the spin-flip
scattering in the high-$B$ limit. Consider a saddle point with zero
energy such that the electric field in its vicinity is $\bbox{\cal
E}=-(V_{xx}x/e)\bbox{\hat{x}}+(V_{yy}y/e)\bbox{\hat{y}}$ with $V_{xx}$
and $V_{yy}$ both positive. Then an electron with the energy $-E<0$
travels from left to right along the $y$ axis.  In the spirit of the
derivation above, it is natural to assume that the main contribution
to $L_{so}^{-1}$ from the saddle points is determined by those of them
that have an anomalously large curvature $V_{yy}$, so that the
equation ${\cal E}_x(y)=0$, ${\cal E}_x(y)$ being taken along the
trajectory with the energy $-E$, is satisfied at anomalously small
$y=i\sqrt{2E/V_{yy}}$. Let $E$ be larger than $2\Delta$, in which case
the saddle point does not split the trajectories with different spins.
Next, to get a feeling for the relevance of the scattering mechanism,
let us assume that the distance between these trajectories with
$|E|\alt\Gamma$ is much smaller than $\lambda$ (i.e., the ratio ${\cal
E}_\Delta/{\cal E}_t$ is not too large) and, besides, put ${\cal
E}_0=0$. Then the matrix element of the spin-flip scattering on
passing the saddle point can be easily evaluated to give
\begin{equation}
  |t_{-+}|^2\propto \exp \left(i {4\Delta\over
      e\lambda^2}\int_{0}^{i\sqrt{2E/V_{yy}}}{dy\over {\cal
      E}_x(y)}\right)=\exp \left( -{2\pi\Delta\over
      \sqrt{V_{xx}V_{yy}}\lambda^2}\right)~.
\end{equation}
Notice that the last expression does not depend on $E$, which is a
peculiar property of the saddle-point potential. The exponent (27) is
typically of order ${\cal E}_\Delta/{\cal E}_t$. To find the
contribution to $L_{so}^{-1}$ from this kind of scattering, we should
average Eq.\ (27) over the parameters $V_{xx}$ and $V_{yy}$.  It is
easy to realize that $V_{xx}$ and $V_{yy}$ are gaussian variables with
zero mean and the correlators $<V_{xx}^2>=<V_{yy}^2>=-\partial^2{\cal
K}_{yy}(0)/\partial y^2$ and $<V_{xx}V_{yy}>=\partial^2{\cal
K}_{xx}(0)/\partial y^2$.  The distribution function of the product
$V_{xx}V_{yy}$ for the saddle-point configuration $V_{xx}>0$,
$V_{yy}>0$ falls off therefore as $\exp (-2V_{xx}V_{yy}d^2/e^2{\cal
E}_t^2)$ (we used the expressions for ${\cal K}_{ii}(y)$ presented
after Eq.\ (20)). Optimization gives for the spin-flip rate near the
saddle points: $L_{so}^{-1}\propto \exp[-\eta' ({\cal E}_\Delta/{\cal
E}_t)^{2/3}]$ with $\eta'=3\pi^{2/3}/2\simeq 3.2$. It follows that the
characteristic curvature of the ``optimum saddle-point" is indeed
anomalously high.  Moreover, it turns out that the resulting
exponential dependence of $L_{so}^{-1}$ on the parameter ${\cal
E}_\Delta/{\cal E}_t$ is similar to that in Eq.\ (25). So we have to
compare the numerical coefficients in the exponents. According to the
estimate given after Eq.\ (25), $\eta<\eta'$. It is this inequality
that justifies the representation of $W_2$ used in Eq.\ (23) and
proves the dominant role of the fluctuations of the form (24).

Yet another essential point is in order. Introducing above the optimum
fluctuation for the case $\Delta\gg\Delta_t$, we performed the
configurational averaging over all possible realizations of the random
potential. This procedure is correct if an electron, traveling along
the classical trajectory, indeed has the possibility to explore the
full spectrum of the fluctuations, i.e., if the length of the
trajectory exceeds $L_{so}$. At given $E$, the classical trajectories
follow equipotential lines, which are closed with the exception of one
percolating equipotential at $E=0$. We are most interested in the
spin-flip rate when the electron energy $E\to 0$, so that the length
\cite{isichenko92} $L(E)\sim d(\Gamma/|E|)^{7/3}$ of the critical
trajectory corresponding to this energy is much larger than $d$.
Nevertheless, it is possible that $L_{so}$ is still larger than
$L(E)$. In that case two electron states which have different spin
projections and are closest to each other in real space are not
resonant any more in the sense that the typical spacing $\delta_E\sim
\hbar v_d/L(E)$ between their energies becomes much larger than the
typical overlap integral $J_{so}$.  Then the relevant characteristic
of the spin-flip scattering is the admixture coefficient
$u_{so}=J_{so}/\delta_E\ll 1$, which describes the hybridization of
the states. Provided ${\cal E}_\Delta\alt {\cal E}_t$, it is given
simply by $u_{so}^2\sim L(E)/L_{so}$. If, however, ${\cal E}_\Delta\gg
{\cal E}_t$ and $u_{so}\ll 1$, the definition of $L_{so}$ used above
becomes meaningless, for typically the electron does not hit the
optimum fluctuation upon ``coming full circle" along the classical
trajectory.  To get $u_{so}$ in this limit, we have to evaluate the
maximum value ${\cal E}_{L(E)}$ of the electric field that the
electron can typically meet on the path $L(E)$. This quantity obeys
the relation $p({\cal E}_{L(E)})L(E)/d\sim 1$, where $p({\cal E})=\exp
[-({\cal E}^2/{\cal E}_t^2)]$ is the probability of finding the
absolute value of the electric field (at a given point) larger than
$\cal E$. The sought function at $L(E)\gg d$ is therefore ${\cal
E}_{L(E)}={\cal E}_t\ln^{1/2}[L(E)/d]$. Thus, if this last expression
is small compared with the amplitude of the electric field in the
optimum fluctuation, which is $\sim {\cal E}_\Delta^{1/3}{\cal
E}_t^{2/3}$ according to Eq.\ (24), the coupling coefficient $u_{so}$
is typically determined by the fluctuation ${\cal E}_{L(E)}$. As a
result, $\ln u_{so}^2 \sim -{\cal E}_\Delta/{\cal E}_{L(E)}$. It
follows that, if ${\cal E}_\Delta\gg {\cal E}_t$, $u_{so}^2$ in the
``mesoscopic" regime is much smaller than $L(E)/L_{so}$.

As argued in Ref.\ 10, the dissipative conductivity $\sigma_{xx}$ at
$\Delta_t\ll\Delta\ll\Gamma$ as a function of the filling factor $\nu$
may exhibit striking behavior in the limit of strong SO coupling.
Namely, provided $T\ll\Delta$ and $L_\phi\ll \xi(0)$, $L_\phi$ and
$\xi(0)$ being the phase-breaking length and the localization length
in the middle between the centers of the Zeeman levels respectively,
$\sigma_{xx}$ has a ``metallic" value $\sim e^2/h$ within an interval
of $\nu$ with well-pronounced boundaries, $\nu_o-\delta\nu <\nu
<\nu_o+\delta\nu$, and sharply falls off outside this interval. Here
$\nu_o$ is an odd integer and $\delta\nu\sim\Delta/\Gamma$. This
boxlike behavior means that the conductivity is high for all energies
lying between the centers of the Zeeman levels and is exponentially
small otherwise. The reason for the ``metallization" of the
conductivity at low $T$ is that the SO coupling changes the nature of
localization: owing to the SO coupling, the localization at
$\nu_o-\delta\nu <\nu <\nu_o+\delta\nu$ is only due to the quantum
interference of multiple-scattered waves, whereas outside this range
it remains classical. Furthermore, when $L_\phi$ exceeds $\xi(0)$ and
two distinct $\sigma_{xx}$-peaks appear, each corresponding to a
different spin projection, the conductivity between the peaks is shown
to fall off with decreasing $T$ only in a power-law manner:
$\sigma_{xx}\sim (e^2/h)\xi(0)/L_\phi$. This ``power-law hopping"
occurs in a wide range of $T$ and goes over into the conventional
variable-range hopping at a very low temperature, which is comparable
with the typical energy spacing on the scale of $\xi(0)$. Thus the
situation is different from that in a white-noise random potential: in
the latter case the variable-range hopping in the QHE regime is likely
to determine the $\sigma_{xx}$-peak width at any
$T$. \cite{polyakov93} Applying now the same reasoning as in Ref.\ 10
to the weak-coupling case $L_{so}\gg L(\Delta)$, we can again reduce
the problem to that of hopping between states which overlap strongly
in real space. The difference is that now the conducting states are
weakly coupled in spin space, which is taken into account by adding
the small factor $u_{so}^2$ (see above) in $\sigma_{xx}$.  Thus, if
$L_\phi\alt\xi(0)$ and $L(\Delta)\alt L_{so}$, $\sigma_{xx}$ between
the peaks is of order $(e^2/h)u_{so}^2$ and does not depend on $T$.
With lowering $T$, when $L_\phi$ becomes larger than $\xi(0)$,
$\sigma_{xx}$ in the middle between the peaks behaves as
$(e^2/h)(\xi(0)/L_\phi)u_{so}^2$.  Eventually, with decreasing
$u_{so}$, a crossover to the picture of two independent
$\sigma_{xx}$-peaks, broadened by the inelastic scattering, takes
place. Notice that the power-law hopping between two adjacent
$\sigma_{xx}$-peaks \cite{polyakov95} should be considered a
characteristic signature of the coupling between the Landau levels
(clearly, it makes no matter whether these are two Zeeman levels,
coupled by the SO interaction, or a number of overlapping Landau
levels with the same spin at low $B$).

It is to the point to remark that, in the case of smooth disorder, the
two spin-split peaks can exhibit a non-trivial temperature dependence
in the limit $u_{so}\to 0$, too. More specifically, their width may be
independent of $T$ in a wide temperature range. This can be seen by
introducing a characteristic temperature $T_s$, at which the
phase-breaking length $L_\phi$ is of the order of $R(\Delta_t)\sim
d(\Gamma/\Delta_t)^{4/3}=d(d/\lambda)^{8/3}$.  Here the length
$R(\Delta_t)$ is the size of the critical percolation cluster with the
energy $\Delta_t$. However, since this energy corresponds to the
crossover between the Anderson-localization regime and that of the
classical percolation, the expression for $R(\Delta_t)$ gives at the
same time the quantum localization length at $|E|\sim \Delta_t$.  If
the inelastic-scattering rate is high enough, such that
$T_s\ll\Delta_t$, the width of the $\sigma_{xx}$-peak in the range
$T_s\ll T\ll\Delta_t$ does not depend on $T$ and is given by
$\Delta\nu\sim \Delta_t/\Gamma$. Indeed, provided $T_s\ll T$, the
phase coherence within the tunneling band $|E|\alt \Delta_t$ is
completely destroyed.  Therefore, $\sigma_{xx}\sim e^2/h$ when the
Fermi level is inside this band. If, however, $T$ is still small
compared to $\Delta_t$, both the hopping transport and the activation
outside the tunneling band yield the conductivity much smaller than
$e^2/h$. As a consequence, the form of the $\sigma_{xx}$-peak is not
affected by the increase of $T$ within the range $T_s\ll T\ll
\Delta_t$.  Note that the presence of several regimes: the
tunneling-dominated regime at $T\alt T_s$, the ``classical" one at
$T\agt \Delta_t$ (with\cite{polyakov93} $\Delta\nu\sim T/\Gamma$),
together with the saturation of $\Delta\nu$ in between (which exists
provided $T_s\ll\Delta_t$), may make it difficult to observe
experimentally the universal critical behavior as $\Delta\nu\to 0$
(see, e.g., Refs.\ 23, 37, and 38).

The preceding analysis of $L_{so}^{-1}$ was restricted to the case
$\Delta_0+\Delta_c\gg \Delta_d$, when the Zeeman splitting is
independent of $\bbox{\cal E}$ and so is spatially homogeneous. In the
opposite limit, the local Zeeman splitting $\Delta(l)=\Delta_d(l)$
follows adiabatically the local electric field. As a result, the
electric field cancels out in the phase $\phi(l)$, which takes the
form $\phi(l)=2(\alpha m/\hbar^2)l$.  Evidently, then, the exponent of
$L_{so}^{-1}$ falls off with increasing ${\cal E}_t$ (see Eq.\ (25))
only as long as $({\cal E}_\Delta/{\cal E}_t)^{2/3}\agt \alpha
md/\hbar^2$, otherwise $L_{so}^{-1}\propto \exp (-4\alpha
md/\hbar^2)$.  This last expression gives the minimum (by modulus)
value the exponent of $L_{so}^{-1}$ can take at given $d$.

So long as the SO-interaction-induced fluctuations of the local
spin-splitting may be neglected, the gap between the energies of two
delocalized states is renormalized by only the constant term
$\Delta_c$ and is given by $2(\Delta_0+\Delta_c)$. Following the model
consideration of Refs.\ 9 and 11, it is tempting to say that when the
random component of the spin-splitting, $\Delta_d(l)$, is dominant,
the energy gap scales as the sample-averaged $<\Delta_d(l)>$. However,
this would not be true; actually, in the classical percolation limit,
the fluctuations of $\Delta_d(l)$ do not lead to any gap at all.  The
point is that, $\Delta_d(l)$ vanishes at the saddle points of the
random potential, whatever the amplitude of the fluctuations.  It is
easiest to see what happens by considering first the periodic
potential $V(\bbox{\rho})\propto\cos(\pi x/d)\cos(\pi y/d)$, which is
a degenerate case in the sense that all saddle points have the same
zero energy. If we neglect the homogeneous splitting
$\Delta_0+\Delta_c$, quasiclassical trajectories at a given energy $E$
follow the equipotentials $V(\bbox{\rho})\pm\Delta_d(\bbox{\rho})=E$.
We observe that the trajectories with opposite spin never coincide
with each other but at $E=0$ they cross at the saddle points of
$V(\bbox{\rho})$. Whether or not a given trajectory passes through the
saddle points is only crucial to the percolation. As a result, in
spite of the fact that the trajectories with opposite spin at $E=0$
are quite different from each other, they both percolate. This same
conclusion holds in the case of a random potential since the only
relevant question then is whether the trajectories with opposite spin
meet at the critical saddle points of the percolation network. They
do, and so the fluctuations of $\Delta_d(\bbox{\rho})$ in between play
no role. Notice that, since $\Delta_d$ at a given point is
proportional to the local electric field, it is not adequate, in the
case of smooth disorder, to consider the fluctuations of
$V(\bbox{\rho})$ and those of the effective magnetic field, induced by
the SO interaction, independently. If they were not correlated with
each other, the percolation transition would split, and the splitting
would scale as $<\Delta_d(\bbox{\rho})>$, which has been demonstrated
by numerical simulation in Ref.\ 11. The above consideration is valid
in the limit of smooth disorder; tunneling at saddle points in the
presence of the SO coupling leads to additional splitting of the
metal-insulator transition\cite{lee94a,wang94} (it is argued in Ref.\
5 that in the case of short-range disorder, similar to the
tunneling-dominated regime, $|E^+_c-E^-_c|\propto
(\alpha^2)^{1/\gamma}$ at $\alpha\to 0$, $\gamma$ being the
localization length exponent).

As the scattering rate at $\Delta\gg \Delta_t$ is strongly suppressed
in the case of smooth disorder, even a weak short-range random
potential, which coexists in reality with the smooth one in any kind
of heterostructure, may provide an essential source of the spin-flip
scattering. Little is known about its strength, so we just write its
correlation function as $<V(0)V(\bbox{\rho})>=w^2\delta(\bbox{\rho})$.
One way to proceed now is to calculate first, by virtue of Eqs.\
(8,9), the contribution to $L_{so}^{-1}$ from the short-range
scattering in the limit ${\cal E}_0\gg {\cal E}_t$. The result is
\begin{equation}
{1\over L_{so}}={1\over \sqrt{2\pi}}\left({\alpha m\over
\hbar^2}\right)^2\left({w\over e{\cal E}_0\lambda^2}\right)^2
\lambda\cos \theta_N {\cal J}_N\left({2\Delta^2\over (e{\cal
E}_0\lambda)^2}\right)~,
\end{equation}
where ${\cal J}_N(t)={2\over \sqrt{\pi}}\int_{t}^{\infty}dt'\exp
(-t')L_N^2(t')(t'-t)^{-1/2}[t'-t(1-1/\cos \theta_N)]$. Then
$L_{so}^{-1}$ in the absence of the homogeneous field ${\cal E}_0$ can
be obtained by simply substituting the absolute value of the smoothly
varying field $\cal E$ for ${\cal E}_0$ in Eq.\ (28) and averaging the
result over $\cal E$ with the distribution function $(2{\cal E}/{\cal
E}_t^2)\exp (-{\cal E}^2/{\cal E}_t^2)$. If $\Delta_d\ll\Delta\ll
\Gamma\lambda/d$, this yields only logarithmic dependence of
$L_{so}^{-1}$ on $\Delta$:
\begin{equation}
{1\over L_{so}}=\sqrt{{2\over\pi}}{\cal J}_N(0)\left({\alpha m\over
\hbar^2}\right)^2\left({w\over e{\cal E}_t
\lambda^2}\right)^2\lambda\ln \left({\Gamma\lambda\over \Delta
d}\right)
\end{equation}
(${\cal J}_0(0)=1$, ${\cal J}_1(0)={7\over 4}$, \ldots ).  This
expression should be compared, if $w$ is known, to Eq.\ (25). In
addition, an inelastic spin-flip scattering may also be important in
the extreme of smooth disorder. Obviously, if $T$ exceeds the energy
$\Delta_i$ of excitations with the in-plane wave vector
$2\Delta/e{\cal E}_t\lambda^2$ (without going here into detail, we
note that, depending on the sample parameters, acoustic phonons or
disorder-induced edge magnetoplasmons, both with a linear spectrum at
low frequencies, may be involved in the transport), it is no longer
necessary to transfer the large momentum to the smooth static
potential, which has been the origin of the exponentially strong
suppression of the purely elastic scattering [Eq.\ (25)]. As for the
temperature behavior of the spin-flip rate at smaller $T$, it depends
on the parameter $T\Delta_i^2/\Delta^3$. If this parameter is small,
the inelastic scattering occurs at $T\ll\Delta_i$ in the regions where
the random static electric field is anomalously high.  This is because
the transition probability in a given place acquires with lowering $T$
an activation factor, the exponent of which, $(\Delta_i/T)({\cal
E}_t/\varepsilon)$, is smaller the higher the local field
$\varepsilon$. Upon optimization with the distribution function of the
electric field, we thus obtain $L_{so}^{-1}\propto \exp
[-3(\Delta_i/2T)^{2/3}]$. The result means that the transitions take
place in the regions with $\varepsilon\sim {\cal
E}_t(\Delta_i/T)^{1/3}\gg {\cal E}_t$ and the characteristic energy
transfer is of order $\Delta_i(T/\Delta_i)^{1/3}$ (which is much
smaller than $\Delta$ according to the above-mentioned condition
$T\Delta_i^2/\Delta^3\ll 1$). One can see that, provided
$T\Delta_i^2/\Delta^3\alt 1$, the exponential factor given by Eq.\
(25) becomes larger than that of the inelastic scattering rate only at
$T$ lower than $\Delta_i\Delta_t/\Delta$. Actually, the elastic
scattering gets dominant at a somewhat higher $T$ due to a difference
in pre-exponential factors. It is straightforward to show that in the
opposite limit, $T\Delta_i^2/\Delta^3\gg 1$, the inelastic transitions
are almost vertical in momentum space and, correspondingly,
$L_{so}^{-1}\propto \exp (-2\Delta/T)$. This last exponent being
compared with that of the elastic scattering loses competition at
$T\sim\Delta^{1/3}\Delta_t^{2/3}$.

\section{Electron-electron interactions}

We have not yet raised the question concerning the influence of
electron-electron interactions on the spin-flip rate. To get an idea
about which modifications should be brought into the above picture
when the Coulomb interaction between electrons is included, let us
suppose the interaction to be weak in the sense that the Bohr radius
$a$ is larger than $\lambda$ (though in the experiments of greatest
interest the ratio of these lengths is of order unity). Let, however,
the fluctuations of the random potential be still so smooth that $d\gg
a$. Then screening of the fluctuations cannot be left out of
consideration. Recall first how matters stand for spinless
electrons. Naturally, the crucial parameter is the ratio $u={\cal
E}_t/(e/\epsilon\lambda^2)$.  One can scan the whole scale of disorder
by changing $u$. As long as $u\ll 1$, the fractional part,
$\tilde\nu$, of the average filling factor $\bar\nu$ is important for
screening. If $u\ll\tilde\nu (1-\tilde\nu )$, the electron
distribution in space is almost homogeneous and the random potential
is almost perfectly screened out \cite{efros88a} (if it were not for
the interaction between electrons, arbitrary weak but smooth disorder
would break up the electron gas into either completely filled or
completely empty regions).  Because of a finite correlation energy,
the concept of compressible quantum Hall liquid in the case of smooth
disorder is compatible with the inevitable existence of a non-zero
electric field within it. In the extreme of high $B$, the
characteristic amplitude of the total self-consistent electrical
potential is of order $e{\cal E}_t\lambda/\sqrt{\tilde\nu (1-\tilde\nu
)}$. \cite{efros88a} However, it is not yet clear exactly in which way
the drift-velocity field produced by this potential yields
localization at all $\tilde\nu$ except $\tilde\nu =1/2$ (ignoring the
self-consistent field within the notion of perfect screening would
lead to the conclusion \cite{efros89} that the dissipative
conductivity is non-zero in a finite range of $\tilde\nu$;
interestingly, the ``restoration" of the conventional one-particle
picture of localization in the limit of weak and smooth disorder may
be solely due to the finite compressibility of the electron liquid,
which, in turn, originates from electron-electron correlations). Thus,
for $\tilde\nu=1/2$ and $u\ll 1$, only one Landau level is responsible
for screening (strictly speaking, electrons in fully occupied levels
also affect the form of the self-consistent potential because of
inhomogeneous mixing of states with different $N$; however, this
polarization effect may be important only in the weak-$B$ limit
\cite{aleiner95}).

If $\tilde\nu (1-\tilde\nu)$ is so small that $\tilde\nu (1-\tilde\nu
)\ll u\ll a/d$, all electrons in the partially filled Landau level
collect together in small droplets near the bottoms of the random
potential. \cite{efros88a} Whatever state of electrons in the droplets
is, the latter are well separated from each other, so that the system
as a whole is in the insulating phase. With further increasing
disorder, $u$ exceeds the ratio $a/d$, which means that the typical
amplitude of the bare random potential becomes larger than the
cyclotron energy.  Provided $u$ is at the same time larger than
$\tilde\nu (1-\tilde\nu)$, electrons in other Landau levels have to
join in participation in screening. Yet the number of participating
levels may be limited to two if $u$ is still small compared with
unity, i.e., $a/d\ll u\ll 1$; then only two adjacent levels can manage
screening. \cite{efros88b} The two-level screening is so effective
that in the high-$B$ limit only a small part of the total area is
occupied by incompressible liquid.  However, this regime obviously
does not appear with increasing $u$ if $\tilde\nu (1-\tilde\nu )\sim
1$.

At still stronger disorder, when $1\ll u$, electrons in the highest
occupied Landau level exhaust their capacity for screening whatever
$\tilde\nu$ is.  Then even the amplitude of the self-consistent
potential becomes larger than $\hbar\omega_c$.  As a result, many
Landau levels come into play and the electron distribution gets close
to that at $B=0$ (similarly to quantum dots/wires without any disorder
\cite{mceuen92,chklovskii92}).  Correspondingly, apart from fine
details, the self-consistent potential may also be obtained within the
framework of the Thomas-Fermi scheme at zero field. Its characteristic
amplitude is then $\sim e{\cal E}_ta$, the ratio of which to
$\hbar\omega_c$ gives the number $\nu_s$ of Landau levels
participating in screening: $\nu_s \sim u\gg 1$.  Lower Landau levels
remain fully occupied.  Clearly, this regime exists only if $1\ll
u\ll\bar\nu$. In the extreme of strong disorder, when $\bar\nu\ll u$,
electrons from all Landau levels condense in droplets lying in the
minima of the random potential. \cite{efros89,shklovskii86} Note that
at $1\ll u$ the self-consistent potential ``pierces" many Landau
levels and so any consideration within a single Landau level becomes
insufficient. The electric field that an electron feels is screened by
all the pierced levels, so that its typical amplitude on the scale of
$d$ is of order ${\cal E}_ta/d$, just as at $B=0$. The Landau level
quantization manifests itself in additional sharp peaks the electric
field exhibits when it breaks through narrow incompressible strips,
which arise whenever one of the levels participating in screening
becomes locally full. \cite{chklovskii92} Provided $u\ll d/a$, the
distance between the incompressible strips is much larger than $a$ and
the characteristic amplitude of the peaks is ${\cal E}_t(a/du)^{1/2}$.
At larger $u$, the oscillations of the electric field on the scale
smaller than $d$ are flattened out and the area occupied by
compressible liquid sharply shrinks. \cite{cooper93,chklovskii93} (It
is not difficult to show that the picture is actually even more
diverse: the above estimate for the crossover point is true if the
cyclotron radius $\lambda\bar\nu^{1/2}\alt a$, which may be
accomplished only for a few lowest Landau levels, otherwise the peaks
of the electric field start to fall down at smaller $u\sim
(d/a)(a^2/\lambda^2\bar\nu)$ because of averaging of the field over
the cyclotron orbit).  The absence of compressible liquid means that
the fluctuations of the filling factor are quantized so that locally
the latter takes only integer values.  The whole picture looks then
much the same as that for non-interacting electrons; the difference is
that the impurity potential is screened (the screening radius being
equal to $a/2$). So we conclude that, once $d$ is supposed to be much
larger than both the Bohr radius and the cyclotron radius, screening
in the quantum Hall regime is important whatever strength of disorder
is (recall that in the case of smooth disorder the quantum Hall
behavior itself occurs unless ${\cal E}_t$ exceeds a critical value of
the order of $e\bar n$, \cite{efros89,shklovskii86} where $\bar n$ is
the average electron concentration; if ${\cal E}_t\gg e\bar n$, the
system resides deep in the insulating phase at any $B$).

One may well ask whether the optimum fluctuation (see above), within
which the electric field is anomalously high, is affected by
screening.  The answer is yes, however large the field is. The point
is that the fluctuation that produces the strong field on the
percolation trajectory consists of a high potential hill followed by a
well of approximately the same depth as the height of the hill, the
percolating trajectory going in the middle between them.  The higher
the electric field, the deeper the well is and the more electrons it
can accomodate. Continuing the same line of argument as in the
consideration of typical fluctuations, we observe that the optimum
fluctuation, being of width $d$, is screened either by one Landau
level or, when the self-consistent potential is larger than
$\hbar\omega_c$, by many Landau levels.  Of course, the hill is
screened only in the linear screening regime (${\cal E}_{opt}\ll e\bar
n$, where ${\cal E}_{opt}$ is the bare value of the field in the
optimum fluctuation).  Otherwise, only the well is screened.  However,
even in the latter case the electric field at the edge of the electron
droplet which fills up the well is much weaker than in the
one-electron scheme.  Indeed, if the screening within the droplet were
perfect, the in-plane electric field would exhibit a square-root
singularity at the edge (see, e.g., Ref.\ 27), being zero inside the
droplet and growing with the distance $x$ from the edge as $\sim {\cal
E}_{opt}(x/d)^{1/2}$ (in the extreme of zero thickness of the layer
occupied by electrons). To get the field at the edge, one should write
$\max (a, \lambda\bar\nu^{1/2})$ in place of $x$. In any case the
resulting field is smaller than ${\cal E}_{opt}$ in the limit at
issue. Hence, ${\cal E}_t$ in the formulae of Sec.\ III has the
meaning of a characteristic amplitude of the self-consistent field.

Let us turn now to the question as to the role of exchange
interactions. The local spin-splitting $2\Delta(l)$ is actually
governed by a competition between three energies: bare Zeeman,
spin-orbit, and exchange. For the quantum Hall liquid, the exchange
effects are of dominant importance in the extreme of weak disorder.
The exchange energy cost of a flipped single-spin excitation in a
homogeneous incompressible liquid at $\lambda\ll a$ is $2\Delta_{ex}=
(\pi/2)^{1/2}(e^2/\epsilon\lambda)$ for $\bar\nu=1$ and scales as
$e^2/\epsilon\lambda\bar\nu^{1/2}$ for higher Landau levels
\cite{ando74} (at $B\to 0$, when $\bar\nu^{1/2}\gg \lambda/a\gg 1$, it
vanishes according to $\Delta_{ex}\sim
(e^2/\epsilon\lambda\bar\nu^{1/2})\ln (\bar\nu^{1/2}a/\lambda)$,
\cite{aleiner95} the additional logarithmic factor being due to
polarization effects).  If $\bar\nu$ is not too large, $\Delta_{ex}$
exceeds greatly both $\Delta_0$ and $\Delta_c$. However, since the
exchange contribution to the effective $g$-factor is controlled by the
local difference between occupation numbers of up-spin and down-spin
states, $\Delta_{ex}$ falls off with increasing disorder. The relevant
parameter is the ratio $\tilde{\cal E}/(e/\epsilon\lambda^2)$, where
$\tilde{\cal E}$ is the characteristic amplitude of the
self-consistent electric field resulting from the above picture of
screening. Specifically, in the extreme of high magnetic field, there
exists a critical value $\tilde{\cal E}_c\sim e/\epsilon\lambda^2$
such that $\Delta_{ex}$ at $\tilde{\cal E}>\tilde{\cal E}_c$ is
strictly zero \cite{dempsey93} (in the sense that, provided the bare
Zeeman and SO contributions are neglected, the separation between the
edge states of opposite spin vanishes). In the critical region of the
low-$\tilde{\cal E}$ phase of this transition a strip of completely
spin-polarized liquid with integer filling appears.  Within the
Hartree-Fock approach, its width $\delta$ scales according to
$\delta\sim\lambda[(\tilde{\cal E}_c-\tilde{\cal E})/\tilde{\cal
E}_c]^{1/2}$. \cite{dempsey93} As the effective confining potential
becomes smoother, $\delta$ gets comparable with $\lambda$ and regions
of fractional filling show up (the essential physics behind this
second transformation is the same as in the collapse of compressible
strips with increasing $\tilde{\cal E}$ for spinless electrons,
\cite{cooper93,chklovskii93} so that it is the exchange energy gain
that outweighs the energy of Coulomb repulsion and thus stabilizes
\cite{dempsey93} the ferromagnetic incompressible state and sharp
edges in a range of $\tilde{\cal E}$). When the compressible phase
appears, the exchange interaction manifests itself in a non-monotonic
dependence of occupation numbers in momentum space \cite{dechamon94}
(even in the framework of the Hartree-Fock scheme), which may be
viewed as a precursor of the formation of wide compressible strips
(note, however, that a similar picture of the sharp-edge
reconstruction with decreasing $\tilde{\cal E}$ has been obtained
without taking exchange into account but with electron-electron
correlations treated within the Hartree approximation for composite
fermions \cite{brey94,chklovskii95}). From the above picture we can
draw the conclusion that even if in a typical place of the electron
trajectory $\tilde{\cal E}\ll e/\epsilon\lambda^2$ and,
correspondingly, $\Delta(l)\sim e^2/\epsilon\lambda$, the
spin-splitting in the optimum fluctuation (see above) still may be
small and be given by the one-particle formula [Eq.\ (4)].  The
electric field $\tilde{\cal E}$ in the optimum fluctuation must then
exceed $e/\epsilon\lambda^2$.

What is possibly the most essential is that the exchange interaction
yields a non-standard mechanism of SO scattering in the quantum Hall
regime.  Since this kind of spin relaxation inevitably involves some
momentum transfer to the random potential (at zero $T$), it is only
natural that, as long as we are concerned with the case of smooth
disorder, $L_{so}$ within the non-interacting theory is exponentially
large. We showed that the scattering rate is in fact much higher than
in the Born approximation; yet, it is out of question to get within
the one-electron picture anything but an exponential dependence of
$L_{so}$ on $d$. Moreover, so far both screening and exchange have
made the spin-flip scattering still more difficult (as they decrease
$\tilde{\cal E}$ and increase $\Delta$). We observe, however, that
their interplay gives rise to phase transitions and, consequently, to
the appearance of new short scales which may ``absorb" the large
momentum.  In the spirit of Ref.\ 28, imagine two edge states with
opposite spin in the effective external field $\tilde{\bbox{\cal E}}$.
Allowing for smooth fluctuations of $\tilde{\cal E}$ along the edges,
consider the neighborhood of the point where $\tilde{\cal E}$ becomes
equal to $\tilde{\cal E}_c\sim e/\epsilon\lambda^2$. Recall that the
characteristic separation of the spin-split edges resulting from the
spontaneous spin-polarization in a clean quantum wire \cite{dempsey93}
is $\lambda$.  Hence, in the case of smoothly varying on this scale
$\tilde{\cal E}$, we can treat the spin-splitting along the edges
adiabatically and introduce the local separation $\delta(l)$. We think
that, in spite of noticeable correlations, the Hartree-Fock
approximation captures the essential physics of the problem and allows
us to write $\delta(l)$ near the transition point $l_c$ in the form
\begin{equation}
\delta(l)=\delta_0(l_c)+\lambda\Theta(l-l_c)\sqrt{(l-l_c)/L_c}~,
\end{equation}
where $\delta_0(l_c)\ll \lambda$, $\Theta(l-l_c)$ is the unit step
function, and $L_c\sim (e/\epsilon\lambda^2)/|\partial\tilde{\cal
E}/\partial l|_{l=l_c}\sim d$ (provided $\tilde {\cal E}(l)$ decreases
from left to right). Then Eq.\ (19) may be used in order to find the
transition probability on passing through the point of phase
separation. Evidently, when $\exp [i\phi(l)]$ rapidly oscillates on
the scale of $d$, most of the contribution to the integral comes from
the singular point of $\delta(l)$.  It is easy to see that if the
characteristic periods of the oscillations near the point $l=l_c$ are
strongly different on different sides of it (which is the case at
$(e^2/\epsilon\lambda\Delta)^3(\lambda/d)\gg 1$), the result is
\begin{equation}
|t_{-+}|^2=\left({\alpha m\over \hbar^2}\right)^2\left({e\tilde{\cal
E}_c \lambda^2\over 2\Delta}\right)^2\sim \left({\alpha\over
a\Delta}\right)^2~.
\end{equation}
Here $2\Delta\ll e/\epsilon\lambda^2$ is the spin-splitting at
$\tilde{\cal E}<\tilde{\cal E}_c$. Remarkably this expression does not
depend on $d$ and falls off with increasing $\Delta$ only in the
power-law manner. Note that Eq.\ (31) implies $\Delta\agt\alpha/a$,
which is fulfilled ``automatically" since $\Delta_d\sim \alpha/a$ in
the electric field of order $e/\epsilon\lambda^2$ [Eq.\ (4)] and,
moreover, in the quantum Hall regime $\Delta_d$ in the field of the
random potential is usually smaller than $\Delta_0$. However,
$\Delta_d$ may be of dominant importance in quantum dots/wires with
strong confinement, in which case the additional small factor
$|<(\chi^+)^*\sigma_x\chi^->|^2=\cos^2\theta$ [Eqs.\ (3,4,17)] should
be introduced in Eq.\ (31). To find $L_{so}^{-1}$ limited by this
mechanism of scattering, the above expression for $|t_{-+}|^2$ should
be multiplied by the linear density $P_c$ (along the electron
trajectory) of the fluctuations with $\tilde{\cal E}=\tilde{\cal
E}_c$: $L_{so}^{-1}=|t_{-+}|^2P_c$.  If we assume that $\tilde{\cal
E}\gg e/\epsilon\lambda^2$ almost everywhere along the edges, then
``bubbles" with large $\delta(l)$ will appear on the electron
trajectories only near the saddle points of the random potential,
which gives $L_{so}^{-1}\sim
d^{-1}(\alpha/a\Delta)^2(e/\epsilon\lambda^2)^2/< \tilde{\cal E}^2>$.
Here $<>$ means averaging along the edge states. In the opposite limit
of weak disorder, most of the area is occupied by inhomogeneous
compressible liquid. In this case, the sharp reconstruction of the
edges occurs only in rare regions where the self-consistent electric
field becomes stronger than $e/\epsilon\lambda^2$. We note that,
though the fluctuations of the bare electric field are gaussian, those
of the self-consistent field, in general, are not. If screening of the
fluctuation with $\tilde{\cal E}=\tilde{\cal E}_c$ were linear, $P_c$
might be written in the form $P_c\sim d^{-1}\exp (-\tilde{\cal
E}_c^2/2< \tilde{\cal E}^2>)$; in fact, the screening radius which
defines the average in the exponent should itself be considered
dependent on the amplitude of the fluctuation (see above). It follows
from the above consideration that in clean quantum wires with weak
disorder there exists an optimum slope of confinement, such that the
electric field at the edge is of the order of $e/\epsilon\lambda^2$,
at which the spin-flip rate reaches maximum with $L_{so}^{-1}\sim
(\alpha/a\Delta)^2d^{-1}$ and falls off if the confinement becomes
smoother or steeper (an order-of-magnitude estimate for $L_{so}$ gives
then $\sim 10 \mu m$). We conclude that, apart from the power-law
dependence on the non-renormalized (by exchange) value of $\Delta$,
$L_{so}^{-1}$ yielded in the high-$B$ limit by this mechanism of
scattering is determined by the distribution of the points on the
electron trajectory where the spontaneous spin-polarization takes
place. One should note, however, that understanding the mechanism of
localization of strongly correlated electrons at smooth disorder will
be necessary in order to make reliable conclusions as to the
experimental situation.

We have already noticed that, when the local $\Delta(l)$ strongly
fluctuates due to the random term $\Delta_d(l)$, $\delta\nu$ (the
difference of the filling factors corresponding to two peaks of
$\sigma_{xx}$) does not scale with the typical amplitude of these
fluctuations. This is because $\delta\nu$ in the extreme of smooth
disorder is only sensitive to the spin-splitting near the saddle
points of the random potential. The drift-velocity-induced splitting
$\Delta_d$ goes to zero at the saddle points and so does not affect
$\delta\nu$, regardless of how large $\Delta_d$ is in a typical place
of the sample. In contrast to this, $\delta\nu$ may be completely
determined by the exchange-induced splitting even though the latter is
small in average. To see how it comes about, suppose that $\Delta(l)$
is solely due to the electron-electron interactions, while the typical
amplitude of the self-consistent electric field $\tilde{\cal E}_t$ is
much higher than $\tilde{\cal E}_c$. Within the mean-field
description, \cite{fogler95} $\delta\nu$ would be strictly zero in
this case. We think, however, that actually $\delta\nu$ for small $N$
is a smooth function of $\tilde{\cal E}_t/\tilde{\cal E}_c$, i.e., it
does not exhibit any critical behavior with changing this
parameter. It is small wonder that the mean-field approach is not
adequate to describe the spin-splitting in the limit of strong
disorder and high $B$ since the problem of finding $\delta\nu$ is then
that of percolation. Indeed, at $\tilde{\cal E}_t\gg \tilde{\cal E}_c$
the local splitting $\Delta(\bbox{\rho})$ vanishes almost
everywhere. Yet, the percolating edge states inevitably pass through
the saddle points of the random potential, near which the electric
field goes to zero and $\Delta(\bbox{\rho})$ is finite. Similarly to
the case of fluctuating $\Delta_d(\bbox{\rho})$ (see above), it is the
behavior of $\Delta(\bbox{\rho})$ near the critical nodes of the
percolating network that is only important to the macroscopic
splitting $\delta\nu$. What is different in comparison with the
SO-induced splitting is that now $\Delta(\bbox{\rho})$ is finite only
close to the saddle points. In order to find $\delta\nu$ we recall
that the ferromagnetic phase appears on the electron trajectory when
$\tilde{\cal E}$ becomes smaller than $\tilde{\cal E}_c \sim
e/\epsilon\lambda^2$. \cite{dempsey93} Since the electric field is a
linear function of the distance $R$ to the saddle point, we get for
the characteristic distance $R_c$ at which the transition occurs
$R_c\sim (e^2/\epsilon\Gamma)(d^2/\lambda^2)\sim
e^2/\epsilon\Delta_t$. At $R<R_c$ the separation $\delta(R)$ between
two trajectories with opposite spin, each corresponding to locally
half-filling of the Zeeman level, appears and starts to grow with
decreasing $R$, first slowly, until it becomes of order $\lambda$ at
$R_c-R\sim R_c$. Then regions of compressible liquid appear on both
sides of the fully polarized strip and grow sharply in width, so that
$\delta$ eventually becomes of order $R_c$. At this point our approach
follows closely that of Ref.\ 32. The question is whether or not the
trajectories with local half-filling of the Zeeman levels will merge
again after they pass through the critical saddle point. According to
the above picture, they will go in opposite directions after the
scattering on the saddle point if $\tilde\nu$ is tuned to be precisely
1 (here the average filling factor $\tilde\nu$ includes both spin
species for given $N$). On the other hand, it is easy to realize that
the trajectories will merge if $|\tilde\nu-1|\gg (R_c/d)^2\sim
(\tilde{\cal E}_c/\tilde{\cal E}_t)^2$. Hence, the up and down spin
trajectories cannot percolate simultaneously, which implies a finite
$\delta\nu$. Clearly, the latter is of order $(\tilde{\cal
E}_c/\tilde{\cal E}_t)^2$. We thereby conclude that
$\delta\nu=F(\tilde{\cal E}_t/\tilde{\cal E}_c)$ is a smooth function,
such that $F(0)=1$ and $F(x)\sim x^{-2}$ at $x\gg 1$. Notice that if
many Landau levels are overlapped, the total filling factor variation
is $\sim\Gamma/\hbar\omega_c$ times larger, $\Gamma$ being the width
of a single Landau level. It should be emphasized also that in the
low-$B$ limit, when the cyclotron radius greatly exceeds $d$, the drop
in $\delta\nu$ with increasing disorder may mimic the mean-field phase
transition, in accordance with Ref.\ 32; but, at any rate, it is
plausible that the triple points \cite{fogler95} do not appear in the
global phase diagram. \cite{kivelson92} Indeed, the order parameter
which governs the spin-splitting of the metal-insulator transition in
the quantum Hall regime is not the sample-averaged spin polarization
$<{\bf S}>$ but, as argued above, rather the typical value of $S^2$
taken at the nodes of the percolation network (of course, this
definition is meaningful only if the quasiclassical limit of smooth
disorder is considered, otherwise a natural generalization is to
consider $<S^2>$ the proper parameter so long as disorder is weak
enough to allow the quantum Hall regime). The simplest case of zero
$<{\bf S}>$ but finite $\delta\nu$ is realized in the model of Ref.\
9: when an effective magnetic field ${\bf B}_e$ coupled to spin in the
one-particle Hamiltonian via the term $\bbox{\sigma}{\bf
B}_e(\bbox{\rho})$ is very strong (in comparison with the scalar
potential) but zero in average. It is the local spin polarization $\bf
S$ that plays the role of ${\bf B}_e$ in the exchange term of the
Hartree-Fock Hamiltonian. Note, however, that a phase diagram with
quadruple points has been suggested recently. \cite{tikofsky95}
Besides, there is experimental evidence \cite{glozman95} for a very
sharp drop of splitting with increasing overlap of the Zeeman levels
at $N=0$.  Possibly relevant to these experimental results is the fact
\cite{lee94a} that the localization length between two Zeeman levels
increases extremely sharply when the SO coupling becomes stronger
(then at given $T$ two $\sigma_{xx}$-peaks should merge rapidly due to
their broadening).

\section{Conclusion}

Our motivation in undertaking this problem stemmed from the
realization \cite{polyakov95} that even a weak SO scattering may
drastically change the picture of QHE. In this paper we presented a
microscopic calculation of the SO scattering length in the case of
smooth disorder. Our main result is that the spin-flip scattering is
strongly inhomogeneous, in the sense that it occurs in optimum
fluctuations of the random potential which appear at rare points of
electron trajectories. The shape of the optimum fluctuations is
characterized by anomalously high electric field. We found a mechanism
of the spin-flip scattering on the optimum fluctuations due to the
exchange-controlled reconstruction \cite{dempsey93} of edge states. We
argue that the spin-splitting of the metal-insulator transition in the
quantum Hall regime is determined also by rare points, -- by nodes of
the percolation network.

\acknowledgements{I am grateful to J. Chalker, J. Hajdu, and M. Raikh
for interesting discussions. I thank also B. Shklovskii for
discussions concerning Ref.\ 32. This work was supported by the
Humboldt Foundation and by the Deutsche Forschungsgemeinschaft through
SFB 341.}

\end{document}